# Generative crystallographic phasing through invariant relationships

Qi Li[1,2]†, Rui Jiao[3,4]†, Liming Wu[5,6,7], Chang Chen[1], Tiannian Zhu[1], Bintang Wang[1], Qiuliang Liu[1], Zhonglong Peng[1], Munan Hao[1], YingPeng Yu[1], Lin Yao[8,9], Wei Ding[1], Mao Su[2], Lei Bai[2], Yang Liu[3,4], Hongming Weng[1], Wenbing Huang[5,6,7]*, Shifeng Jin[1]*, Xiaolong Chen[1]*

[1]The Beijing National Laboratory for Condensed Matter Physics, Institute of Physics, Chinese Academy of Sciences, Beijing, 100190, China.
[2]Shanghai Artificial Intelligence Laboratory, China.
[3]Dept. of Comp. Sci. and Tech., Institute for AI, Tsinghua University, Beijing, China.
[4]Institute for AIR, Tsinghua University, Beijing, 100084, China.
[5]Gaoling School of Artificial Intelligence, Renmin University of China, Beijing, 100872, China.
[6]Engineering Research Center of Next-Generation Intelligent Search and Recommendation, MOE, China.
[7] Beijing Key Laboratory of Big Data Management and Analysis Methods, China.
[8] School of Computer Science, Shanghai Jiao Tong University, Shanghai, 200240, China
[9] Zhongguancun Academy, Beijing, 100097, China
†These authors contributed equally: Qi Li, Rui Jiao.
*Corresponding authors. Email: hwenbing@ruc.edu.cn; shifengjin@iphy.ac.cn; chenx29@iphy.ac.cn

**Crystal structure determination requires the phases of scattered waves — yet diffraction measures only their intensities. Direct methods exploit phase invariants but become less reliable as diffraction information diminishes. Learned phase prediction has lowered the resolution barrier, yet remains primarily confined to centrosymmetric crystals with binary phases. We introduce PhiGen, a generative reformulation of traditional direct methods that learns origin-independent phase relationships for binary and continuous phasing. Across 210 space groups, including groups absent from training, it recovered high-quality maps for 99.0% of centrosymmetric structures and invariant-consistent phases for 92.8% of non-centrosymmetric structures. From simulated 3 Å zeolite powder data, the network recovered framework maps for 84.2% of held-out structures, versus 1.0% for Superflip. For experimental ZSM-25 and TNU-9, generated phases seeded high-resolution phase extension. These results suggest a route to structure determination from low-resolution, incomplete, and overlapped diffraction data.**

For over a century, crystallography has revealed the atomic architecture of matter — from rock salt (*1*) to the ribosome (*2*, *3*) — with a precision unmatched by other structural technique. Yet crystallography experiment must overcome a fundamental obstacle: detectors record the intensities of scattered X-rays, electrons, or neutrons, from which the amplitudes of the diffracted waves are derived, but not the phases required to reconstruct the crystal structure by Fourier synthesis. This missing information is the crystallographic phase problem (*4*). Recovering the phases is possible because reflections produced by the same atomic structure are not independent. Their relationships carry information that no individual intensity measurement alone contains.

Classical direct methods—recognized by the 1985 Nobel Prize in Chemistry—turn these relationships into a principled approach to phase determination (*5–11*). Individual phases change when the arbitrary origin of the unit cell is translated, but certain combinations, known as structure invariants, remain unchanged (Fig. 1A-C). Atomicity gives rise to amplitude-dependent statistics of these invariants, while crystallographic symmetry imposes exact relations among symmetry-equivalent reflections (Fig. 1D). Together, these statistical and algebraic relationships underpin phase determination for both centrosymmetric and non-centrosymmetric crystals (*12–14*). However, as observations are lost or made uncertain, symmetry relations remain exact, while the statistical inference drawn from them becomes less reliable (*15*, *16*). Resolution truncation blurs atomic detail; incomplete angular coverage removes reflections that would otherwise constrain the solution; powder diffraction merges reflections with similar spacings into overlapping peaks (*17*, *18*). Under such conditions, conventional phasing can falter—particularly for weakly scattering crystals and complex framework materials, where structure determination has sometimes relied on auxiliary chemical information, specialized structural models, or electron microscopy (*19*, *20*).

The analytical probability relations of direct methods are approximations to distributions that are, in principle, learnable from structural data. This insight has reopened the question of whether phases can be recovered from information-poor diffraction data. Neural networks have been applied to direct phase prediction (*21–24*), phase seeding for dual-space refinement (*25–27*), and generative structure solution from powder patterns (*28–30*).

The landmark PhAI study showed that a network trained on millions of artificial structures could recover phases from 2 Å data using substantially less diffraction information than conventional direct methods, establishing that atomic resolution is not a strict requirement for learned phasing. The advances of PhAI and its subsequent graph-based extensions, however, were principally developed in the centrosymmetric binary setting, where phases can be represented as 0 or π after an appropriate origin choice (*23*, *24*). Extending learned phasing to non-centrosymmetric crystals presents a central additional challenge (*25*): phases are continuous, and the phase of an individual reflection depends on the arbitrary choice of unit-cell origin, of which there are infinitely many. The physically meaningful information instead resides in coupled invariant combinations of phases. This motivates a formulation that generates complete, coupled phase sets while respecting the invariant relationships at the heart of classical direct methods.

Here we introduce PhiGen, a generative reformulation of traditional direct methods that preserves their exact crystallographic relationships while learning the statistical inference required for phase recovery. By organizing inference around triplet invariants and structure-factor algebra, PhiGen extends learned phasing beyond binary centrosymmetric phases to continuous phases in non-centrosymmetric crystals. We evaluate this framework across 210 space groups, including symmetries absent from training, and under progressively information-poor conditions involving resolution loss, incomplete and restricted sampling, and powder reflection overlap. In experimental ZSM-25 and TNU-9 powder data, PhiGen-generated low-resolution phases provide seeds for higher-resolution phase extension. The results establish invariant-guided generation as a route to coupled phase recovery across symmetry classes and to structure solution from challenging diffraction data.

## PhiGen generates phase sets through invariant relationships

Direct methods exploit relationships among the phases of reflections produced by the same atomic structure. Although the phase of an individual reflection depends on the arbitrary choice of unit-cell origin, certain combinations do not. For a triplet satisfying $\boldsymbol{h} + \boldsymbol{k} = \boldsymbol{l}$, the phase combination

$$\Phi_{hkl} = \varphi(\boldsymbol{h}) + \varphi(\boldsymbol{k}) - \varphi(\boldsymbol{l}) \quad (1)$$

is invariant to an origin translation. The origin-dependent phase shifts cancel because $h + k - l = 0$. Triplet invariants therefore provide origin-independent relationships among the missing phases, even though the individual phases themselves are not uniquely defined.

Measured amplitudes provide statistical information about the values of these invariants. Classical direct methods derive this information from statistical models of atomic arrangements and normalized structure factors. In a representative triplet approximation,

$$p(\Phi_{\boldsymbol{hkl}} \mid |E_{\boldsymbol{h}}|, |E_{\boldsymbol{k}}|, |E_{\boldsymbol{l}}|) \propto \exp[\kappa_{\boldsymbol{hkl}} \cos(\Phi_{\boldsymbol{hkl}})] \quad (2)$$

where $E$ denotes a normalized structure factor and $\kappa$ depends on the corresponding normalized amplitude product and structural statistics (*11*). A larger $\kappa$ gives a more concentrated and informative phase relationship. Overlapping triplets then form a coupled network that can be used to refine and extend phase estimates. Direct methods combine these statistical relationships with exact constraints imposed by space-group symmetry. A symmetry operation with rotational part $\boldsymbol{R}_s$ and translational part $\boldsymbol{T}_s$ relates symmetry-equivalent structure factors according to

$$F(\boldsymbol{h}\boldsymbol{R}_s) = F(\boldsymbol{h}) \exp(-2\pi i\, \boldsymbol{h} \cdot \boldsymbol{T}_s) \quad (3)$$

using the row-vector reciprocal-space convention (*12*). Thus, symmetry-equivalent reflections have equal amplitudes and deterministically related phases. Symmetry reduces the number of independent phase variables, whereas triplet statistics provide information about their likely values.

PhiGen retains both ingredients of direct methods while replacing their analytical statistical model with learned, amplitude-conditioned generative inference (Fig. 2). Measured amplitudes remain fixed, whereas the unknown phases are generated jointly. Each independent reflection is represented as a node, and each valid relation $h + k = l$ defines a triplet hyperedge. TripletsNet combines the measured amplitudes, reciprocal-lattice geometry, current phase state, triplet relationships, and supplied space-group operations to predict phase updates. PhiGen generates phases only for symmetry-independent reflections and reconstructs the equivalent phases algebraically. Iterative message passing through overlapping triplets allows information to propagate across the coupled phase set, rather than assigning phases independently reflection by reflection.

The phase domain determines the generative procedure. For centrosymmetric crystals, independent phases are restricted to 0 or $\pi$, and PhiGen uses "discrete diffusion" (*31*). For non-centrosymmetric crystals, phases are continuous angles on the circle $S^1$, and PhiGen uses "circular flow matching" (*32*, *33*) without artificially discretizing the phase domain. In both branches, generation begins from a random phase state and iteratively updates the phases while keeping the measured amplitudes fixed. The generated phase set is combined with the measured amplitudes and reconstructed symmetry equivalents for Fourier synthesis.

PhiGen therefore imposes the crystallographic structure of the inference problem—triplet closure and space-group algebra—while learning the amplitude-conditioned statistical inference needed to select useful coupled phase sets. It extends the invariant framework of direct methods from analytical phase estimation to generative inference, and from binary centrosymmetric phases to continuous non-centrosymmetric phases.

## PhiGen generalizes invariant phasing to continuous phases and unseen symmetries

We tested whether PhiGen's invariant-guided formulation supports phase recovery across diverse crystallographic symmetries, extends to continuous non-centrosymmetric phases, and transfers to space groups absent from training. PhiGen was trained on curated Alexandria structures (*34*, *35*) comprising 51,497 centrosymmetric structures across 71 space groups and 50,530 non-centrosymmetric structures across 92 space groups. The held-out benchmark contained 4,047 structures across 210 space groups: 1,963 centrosymmetric structures in 88 groups and 2,084 non-centrosymmetric structures in 122 groups. We used complete, calculated 1 Å diffraction data and supplied the correct space-group operations to all methods, including for groups absent from PhiGen's training set. This setting isolates phase inference from experimental noise, incomplete measurements, and incorrect symmetry assignment. We compared PhiGen with two established phasing programs:

Superflip (*36*), a widely used dual-space method (*37*, *38*), and SHELXS (*39*), a classical direct-methods (*11*) program that implements the analytical approach PhiGen generalizes.

Centrosymmetric recovery was evaluated by the correlation coefficient (CC) between electron-density maps reconstructed from predicted and reference phases using common amplitudes (higher is better). Non-centrosymmetric recovery was evaluated by invariant consistency (IC), the amplitude-product-weighted mean absolute circular error between predicted and reference triplet invariants (lower is better; full definitions in the supplementary information). PhiGen results report the highest CC or lowest IC among 10 independently generated candidates, selected using the reference structure. This reference-selected, best-of-10 protocol represents a best-case evaluation of PhiGen, whereas Superflip and SHELXS results reflect their internal selection criteria, with 10,000 random searches for SHELXS and 20 random trials for Superflip.

The centrosymmetric benchmark tested whether the invariant-guided generative formulation could recover binary phases across 88 space groups. With the correct space-group constraints supplied, PhiGen recovered maps with CC $\geq$ 0.95 for 99.0% of structures and met the corresponding mean-CC criterion in all 88 tested space groups, including all 17 groups absent from training (Fig. 3A). Superflip and SHELXS met the same group-level criterion in 67 and 25 groups, respectively. These results demonstrate broad centrosymmetric phase recovery of PhiGen, including in space groups absent from training.

The non-centrosymmetric benchmark tested the extension from binary phase assignments to continuous, origin-dependent phase spaces. Here, phases are continuous, and their absolute values change with the arbitrary choice of unit-cell origin. We therefore used triplet-invariant agreement to assess phase recovery without explicit origin alignment. PhiGen achieved a median IC of 0.054, with 92.8% of structures meeting the invariant-consistency criterion IC $\leq$ 0.20, compared with 72.9% for Superflip and 24.1% for SHELXS (Fig. 3B). At the space-group level, PhiGen met the mean-IC criterion in 115 of 122 groups, whereas Superflip and SHELXS met it in 75 and 15 groups, respectively. These results establish that the generative formulation recovers continuous invariant relationships across diverse non-centrosymmetric space groups.

Space groups absent from training provided a direct test of crystallographic transfer. In centrosymmetric groups, PhiGen showed little loss from seen to unseen symmetries: CC $\geqslant$ 0.95 was achieved for 99.2% and 97.9% of structures, respectively. Unseen non-centrosymmetric groups were more demanding, but PhiGen generated candidates meeting the IC $\leqslant$ 0.20 criterion for 85.2% of structures, compared with 74.6% for Superflip and 26.2% for SHELXS on the same dataset. Figure 3C connects invariant agreement to real-space reconstruction: for an unseen chiral structure, PhiGen produced a more localized density map with IC = 0.099, compared with 0.595 for Superflip and 1.751 for SHELXS. Across triplet strengths, the invariant phases predicted by PhiGen closely followed the reference values (Fig. 3D), confirming that the recovered phases satisfy the underlying invariant constraints.

Finally, we tested phase recovery without supplying the full space-group symmetry. Centrosymmetric structures were represented in *P*-1, retaining inversion symmetry, and non-centrosymmetric structures in *P*1 (supplementary materials). Performance decreased for all methods, but PhiGen retained the strongest recovery. For centrosymmetric structures, PhiGen recovered maps with CC $\geqslant$ 0.95 for 78.6% of structures, compared with 67.0% for Superflip and 38.2% for SHELXS. For non-centrosymmetric structures, 90.3% of PhiGen cases met the IC criterion, compared with 57.3% for Superflip and 0% for SHELXS. Explicit symmetry information thus improves recovery, while learned invariant-based inference remains effective when nontrivial symmetry operations are withheld.

## PhiGen recovers phases from resolution-limited and incomplete diffraction data

Classical ab initio phasing is most reliable when diffraction data reach atomic resolution, are highly complete, and provide redundant relationships among reflections. We therefore tested whether PhiGen could retain phase information when these supports were deliberately weakened. Resolution truncation obscures atomic detail; loss of weak reflections removes observations and associated triplet relationships; angular restriction leaves a coherent region of reciprocal space unobserved. Limited resolution and incomplete coverage are important obstacles in diffraction from small or fragile crystals, including limited-tilt electron diffraction experiments (*40*).

To separate information loss from space-group variation, we constructed a controlled benchmark from artificial organic structures in space group $P2_1/c$, following previous work on synthetic crystallographic training data (*23*, *41*). A separate PhiGen model was trained for each degradation condition on approximately 70,000 structures, and 1,150 held-out structures were evaluated against Superflip and SHELXS. CC was calculated using only the reflections retained under each condition. Complete 1 Å data served as a positive control: all three methods achieved CC > 0.8 for more than 97% of structures (Fig. 4A). These results establish a baseline: under favorable resolution and completeness, conventional phasing handles these structures effectively.

Resolution truncation tested recovery at the practical resolution boundary established by earlier learned phasing (*23*). At 2 Å resolution, PhiGen achieved CC > 0.8 for 85.7% of structures, compared with 10.5% for Superflip and 25.8% for SHELXS (Fig. 4B). The median CC remained 0.990 for PhiGen, whereas the medians for Superflip and SHELXS fell to 0.385 and 0.549. Thus, even when resolution truncation obscured atomic detail, PhiGen retained accurate phase information for the remaining reflections across most of the benchmark.

Retention of only the strongest 10% of reflections probed a different requirement of conventional phasing: redundancy among reciprocal-space observations. Motivated by weakly scattering crystals, this controlled selection removes most reflections and many of the triplet relationships available in complete data. PhiGen nevertheless achieved CC > 0.8 for 88.5% of structures and retained a median CC of 1.000 over the observed subset. Superflip and SHELXS reached the same threshold for 14.4% and 30.1% of structures, with median CC values of 0.341 and 0.480, respectively (Fig. 4C).

Angular restriction tested recovery under anisotropic reciprocal-space coverage. We retained reflections within a cone centered on the reciprocal [001] direction, with a half-angle of 50°, corresponding to a full opening angle of 100°. PhiGen achieved CC > 0.8 for 76.7% of structures with a mean CC of 0.882 and a median of 0.996, consistent with the loss of an entire region of reciprocal space rather than a uniform thinning of observations. Superflip and SHELXS reached the same threshold for 18.4% and 20.6% of structures, respectively (Fig. 4D). In paired comparisons, PhiGen achieved higher CC than Superflip and SHELXS for 94.8% and 80.7% of structures, respectively. The advantage therefore extended across the benchmark rather than being confined to a small set of favorable cases.

A representative structure illustrates the practical consequence of information loss. All three methods recovered localized density from complete 1 Å data. Under 2 Å resolution, strongest-10% retention, and angular restriction (see inserts in Fig. 4), PhiGen retained density associated with the reference atomic arrangement, with CC values of 0.996, 1.000, and 0.899, respectively. Superflip and SHELXS produced less complete or more fragmented maps (Fig. 4A–D). Together, these controlled tests show that condition-specific learning enables the invariant-guided framework to recover useful phases even when high-resolution observations, most weak reflections, or broad angular coverage are unavailable.

## PhiGen recovers zeolite frameworks from overlapped powder diffraction

Powder diffraction compounds the forms of information loss examined above by projecting three-dimensional reciprocal space onto a one-dimensional profile, causing multiple reflections to share or nearly share the same measured intensity. This loss of reflection separability is particularly severe for large framework structures, in which dense diffraction patterns and limited resolution can prevent conventional methods from obtaining an initial phase set. We therefore used zeolite powder diffraction as a stringent test of whether invariant-guided generation can recover useful phase information from severely overlapped observations and supply starting phases for conventional high-resolution extension.

PhiGen was trained on 48,533 simulated 3 Å zeolite powder datasets from the Predicted Crystallography Open Database and the International Zeolite Association database (*42–44*), spanning 63 space groups, and evaluated on 1,854 held-out structures spanning 87 space groups. The simulated inputs incorporated reflection overlap rather than treating the observations as independent single-crystal amplitudes. We used CC $\geq$ 0.7 as an operational threshold for framework-level recovery and CC ≥ 0.8 as a stricter map-recovery criterion. Across the benchmark, PhiGen achieved a mean CC of 0.838 and a median CC of 0.869, recovering framework-level maps for 84.2% of structures, compared with 1.0% for Superflip. Performance remained robust as framework size increased: in the 20,000–30,000 Å³ volume bin, which contained only space groups absent from training, PhiGen achieved a mean CC of 0.757 and recovered 64.5% of structures, whereas Superflip recovered none (Fig. 5B).

We next examined ZSM-25 and TNU-9, two historically difficult zeolite structures whose large, complex frameworks resisted routine ab initio phasing and required auxiliary phase information for their original determinations (*19*, *45*). ZSM-25 is a cubic zeolite with 16 crystallographically independent tetrahedral sites and a unit-cell volume of approximately 91,536 Å³ — more than 4,500 framework atoms in all. TNU-9 contains 24 crystallographically independent tetrahedral sites. Recent phase-seeding work (*25–27*) has shown that approximate phases for a limited subset of reflections can guide established dual-space methods toward complete solutions. We therefore tested whether PhiGen-generated low-resolution phases could supply such seeds.

We first tested this strategy under controlled conditions using calculated ZSM-25 powder data. Four independently generated PhiGen outputs were supplied to Superflip as 3 Å phase seeds. Three PhiGen output had high initial CC values of 0.780 to 0.830, and subsequent phase extension converged to high-quality solutions with final CC values of 0.957 to 0.961; the remaining output had an initial CC of 0.060 and did not converge to the correct solution. Thus, the high-quality PhiGen outputs consistently supplied sufficient phase information for Superflip to complete the high-resolution phase set (Fig. 5C and supplementary materials).

We then applied this workflow to experimental ZSM-25 powder data. Electron-beam damage limited the resolution of the rotation-electron-diffraction data used in the original determination and prevented direct-method initialization (*45*), even though the reflections were individually resolved rather than merged by powder overlap. The original solution instead transferred the phases of 21 strong reflections from corresponding reflections of the related PAU framework, allowing all 16 independent T sites to be located. We performed structure-free Le Bail fitting (*46*) of the as-made ZSM-25 powder pattern and extracted structure-factor estimates to 1.3 Å; 98.6% of the reflections belonged to overlap groups. From the data truncated at approximately 3 Å, one of four PhiGen outputs generated 181 phases with CC = 0.707. Using these phases to seed extension against the 1.3 Å data yielded a final map with CC = 0.913, resolving all 16 independent T sites and 80% of framework oxygen sites (Fig. 5C). In contrast, four matched Superflip-generated 3 Å seeds produced no solved extensions, with a best final phase CC of 0.130. For each experimental phase extension run, the reported endpoint was selected using Superflip's internal criterion, whereas agreement with the reference structure was assessed retrospectively.

As an independent experimental test, we applied the same strategy to TNU-9, whose original structure determination combined powder diffraction with electron-microscopy-derived phase information (*19*). One of 40 PhiGen outputs supplied 219 phases with a 3 Å CC of 0.786; at this resolution the framework topology was already discernible. Extension against experimental data to 1.16 Å produced a framework map with CC = 0.755 in which individual framework atoms were resolved; subsequent electron-density polishing raised the correlation to CC = 0.80. In contrast, none of 40 independently generated Superflip phase seeds produced a solved extension; the best final CC was 0.322 (Fig. 5D).

## Conclusion

PhiGen reformulates crystallographic phasing as the generation of mutually consistent phase sets governed by origin-independent invariants rather than the prediction of individual phases. This generative generalization of direct methods extends learned phasing from binary centrosymmetric phases to continuous non-centrosymmetric ones. PhiGen transferred across 210 space groups; condition-specific models further recovered phases from low-resolution, incomplete, wedge-limited, and overlapped data that challenged SHELXS and Superflip. Recovery of zeolite frameworks from simulated powder data and experimental ZSM-25 and TNU-9 shows that invariant-guided generation can provide useful phase seeds even when reflection overlap obscures individual intensities.

The current models remain specialized by data regime and triplet-selection strategy, and the powder simulations do not capture all experimental effects. Future work should accommodate variable data quality, disorder, preferred orientation, systematic intensity errors, and uncertainty in generated phase sets. PhiGen is therefore not a replacement for crystallographic refinement or validation, but a phasing engine that supplies invariant-consistent hypotheses when conventional procedures cannot initialize. In that role, invariant-constrained generative phasing could open structure determination to sparse, continuous, and degenerate diffraction data that have historically resisted ab initio initialization.

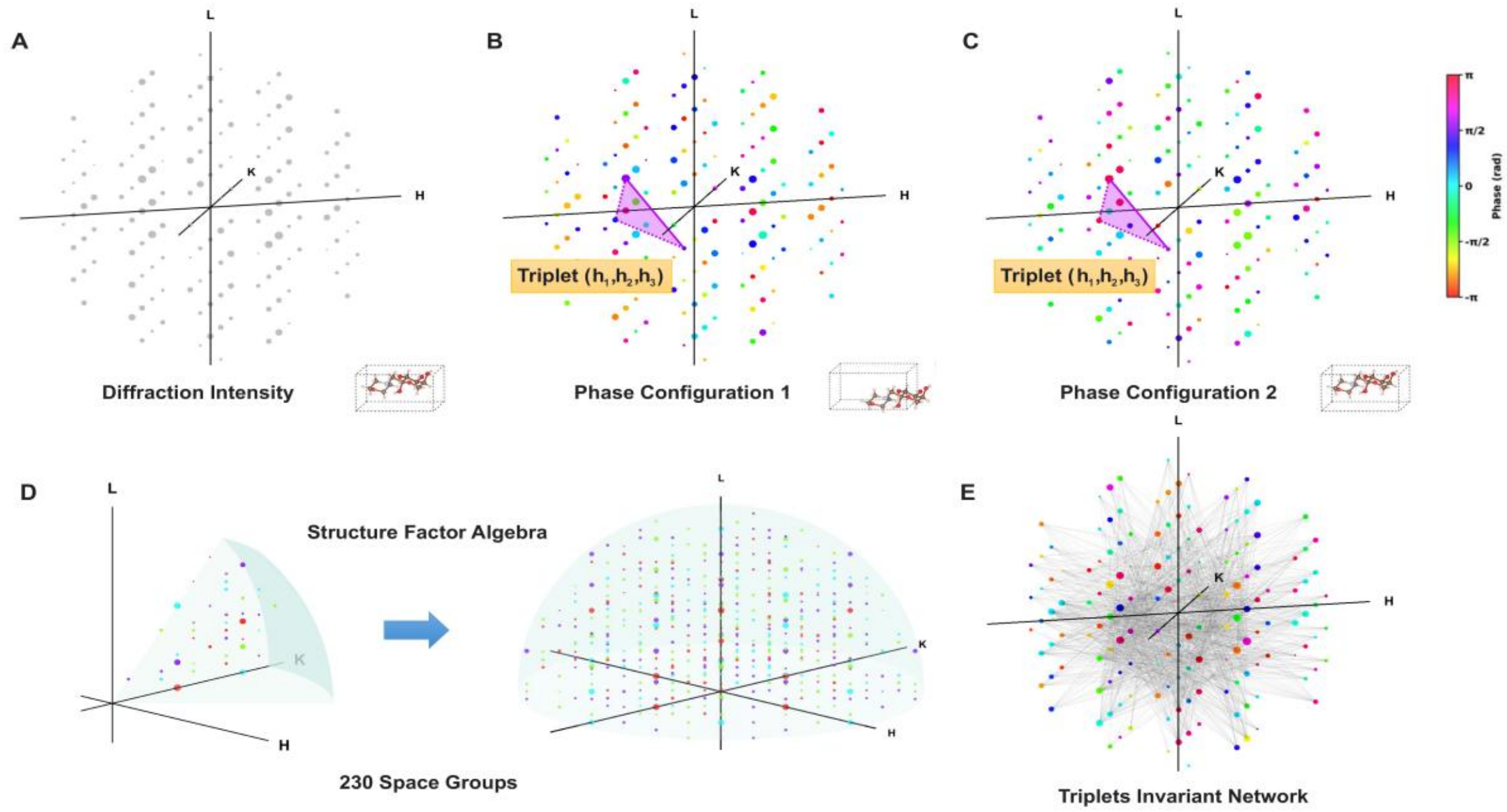

**Figure 1. Invariant relationships make crystallographic phasing possible.** (A) Diffraction data provide structure-factor amplitudes $|F(\boldsymbol{H})|$ in reciprocal space, whereas the phases required for electron-density reconstruction are unobserved. (B and C) Two phase representations of the same structure related by an origin shift assign different phases to individual reflections but preserve the highlighted triplet invariant $\Phi = \varphi(\boldsymbol{H}) + \varphi(\boldsymbol{K}) - \varphi(\boldsymbol{L})$, where $\boldsymbol{H} + \boldsymbol{K} = \boldsymbol{L}$. (D) Space-group operations generate symmetry-related reflection instances from the independent set, the related phases are constrained by structure factor algebra. (E) Overlapping triplets connect reflections into a redundant network of mutually reinforcing constraints. Classical direct methods exploit this network to identify phase sets consistent with the measured amplitudes.

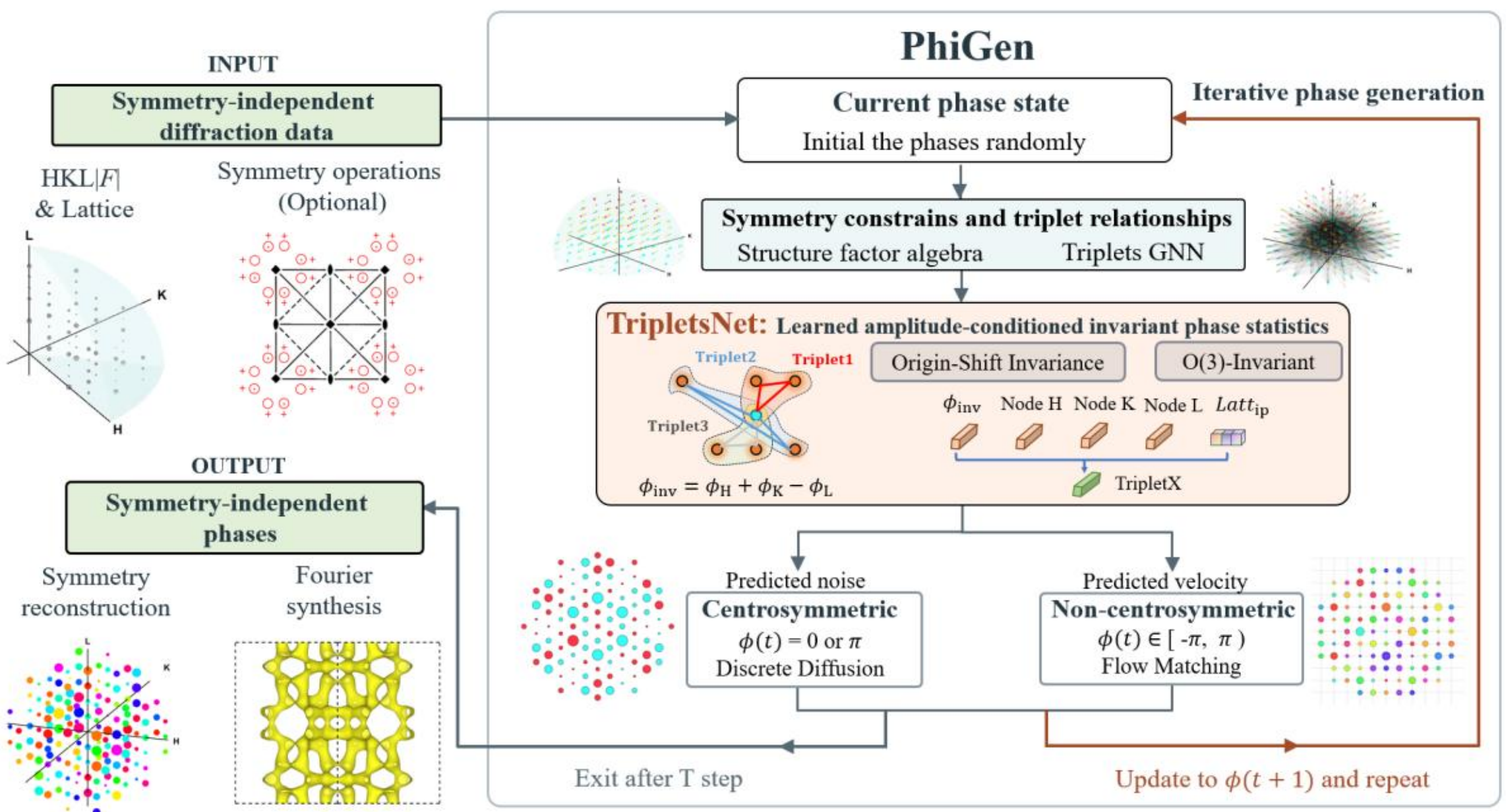

**Figure 2. PhiGen performs generative phasing through a symmetry-expanded triplet graph.** Measured amplitudes, independent Miller indices, and space-group symmetry operations are used to construct a reciprocal-space graph of symmetry-related reflection instances and triplet closure relations. TripletsNet propagates information through this graph while updating only the independent phase variables. A discrete diffusion branch generates binary centrosymmetric phases, whereas circular flow matching transports continuous non-centrosymmetric phases on $[-\pi, \pi)$. The final symmetry-equivalent phases are reconstructed algebraically and combined with the measured amplitudes for Fourier synthesis.

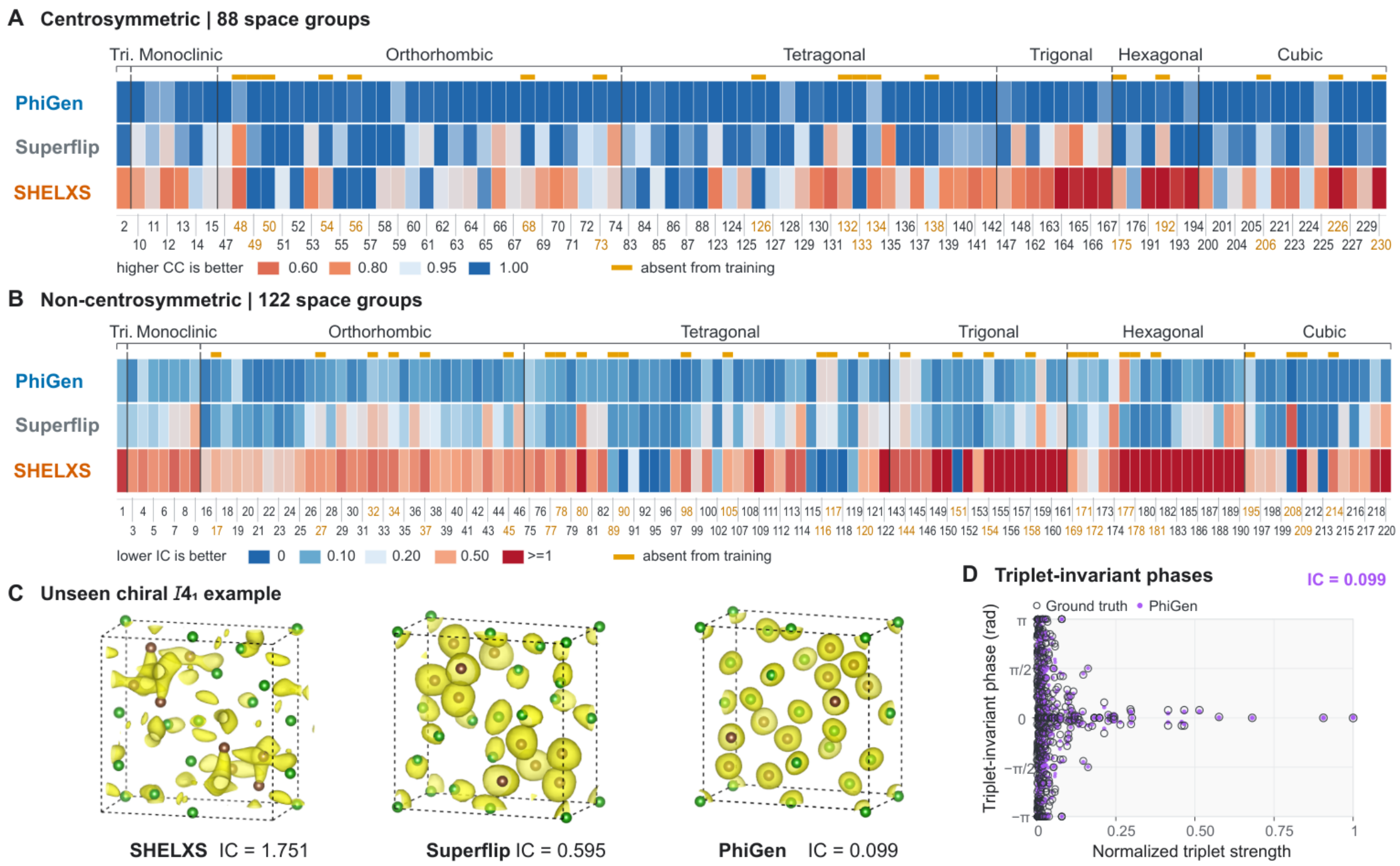

**Figure 3. PhiGen generalizes invariant phasing to continuous phases and unseen symmetries.** (A) Centrosymmetric benchmark comprising 1,963 held-out structures in 88 space groups, grouped by crystal system. Each column reports the mean electron-density correlation coefficient (CC; higher is better) for PhiGen, Superflip, and SHELXS; orange marks identify space groups absent from training. (B) Non-centrosymmetric benchmark comprising 2,084 structures in 122 space groups, evaluated by origin-independent invariant consistency (IC; lower is better). The values at right give the fraction with IC $\leq$ 0.20 and the number of groups passing this threshold by mean IC. (C) Density maps for a representative unseen chiral $I4_1$ structure. PhiGen produced the most localized map and the lowest IC (0.099), compared with Superflip (0.595) and SHELXS (1.751); spheres show the reference atomic positions. Maps were origin-aligned to the reference before visualization. (D) Reference and PhiGen triplet-invariant phases plotted against normalized triplet strength for the same structure. Open black and filled purple symbols denote reference and PhiGen values, respectively. A space group was considered unseen when no structures from that group appeared in training.

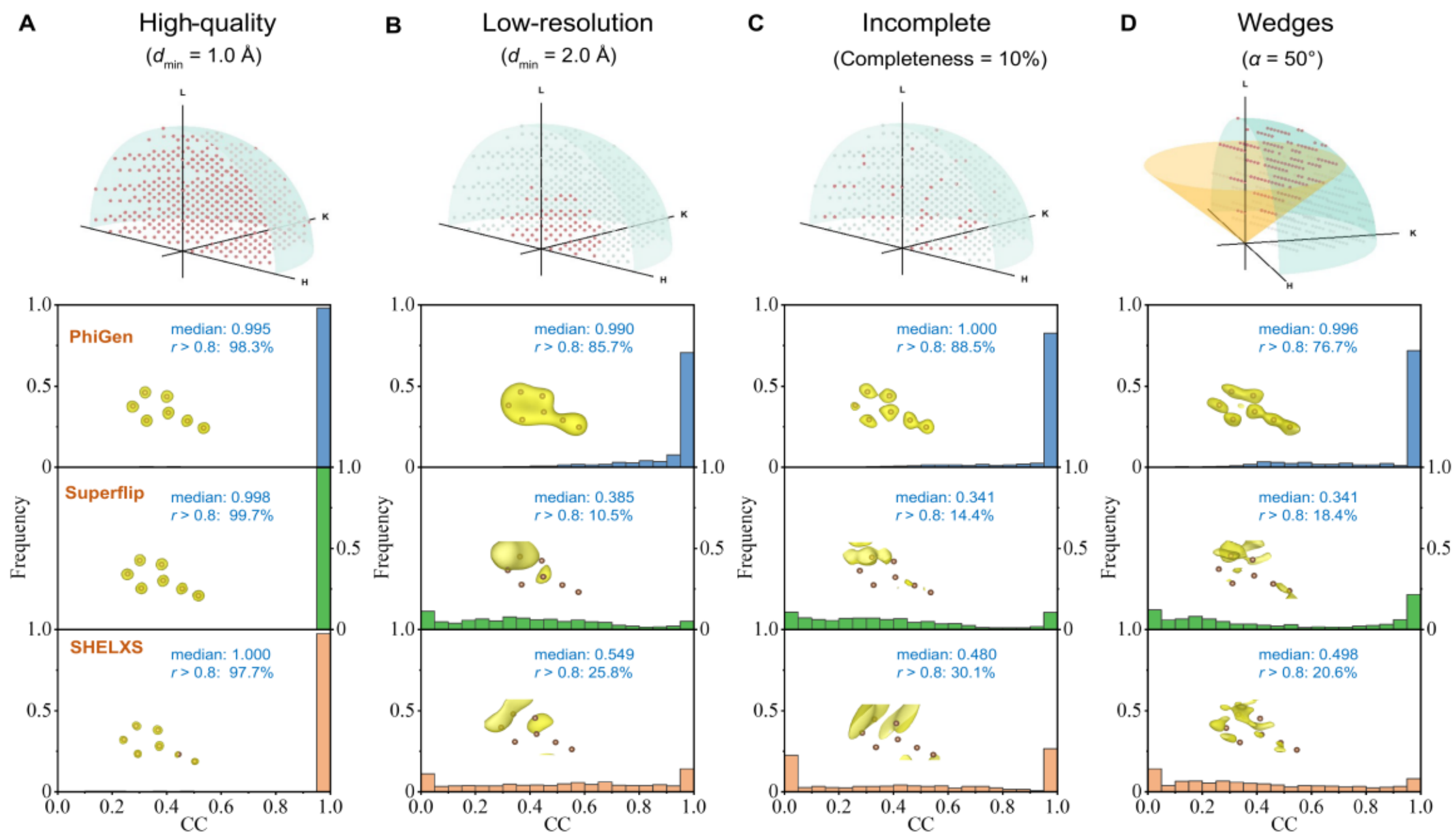

**Figure 4. PhiGen recovers phases from resolution-limited and incomplete diffraction data.** Matched benchmark of 1,150 held-out $P2_1/c$ structures; a separate PhiGen model was trained for each degraded-data condition. Each panel shows the retained reciprocal-space information, electron-density CC distributions for PhiGen, Superflip, and SHELXS, the median CC and fraction with CC ≥ 0.8, and a representative density-map inset. (A) Complete 1 Å diffraction data provide a positive control: the three methods recover 97.7 to 99.7% of structures. (B) At 2 Å resolution, PhiGen retains a median CC of 0.990 and practical recovery for 85.7% of structures, compared with 0.385 and 10.5% for Superflip and 0.549 and 25.8% for SHELXS. (C) When only the strongest 10% of reflections are retained, PhiGen retains a median CC of 1.000 and recovers 88.5%, compared with 0.341 and 14.4% for Superflip and 0.480 and 30.1% for SHELXS. (D) When reflections are restricted to a 50° angular wedge around a selected reciprocal-space axis, PhiGen retains a median CC of 0.996 and recovers 76.7%, compared with 0.342 and 18.4% for Superflip and 0.498 and 20.6% for SHELXS.

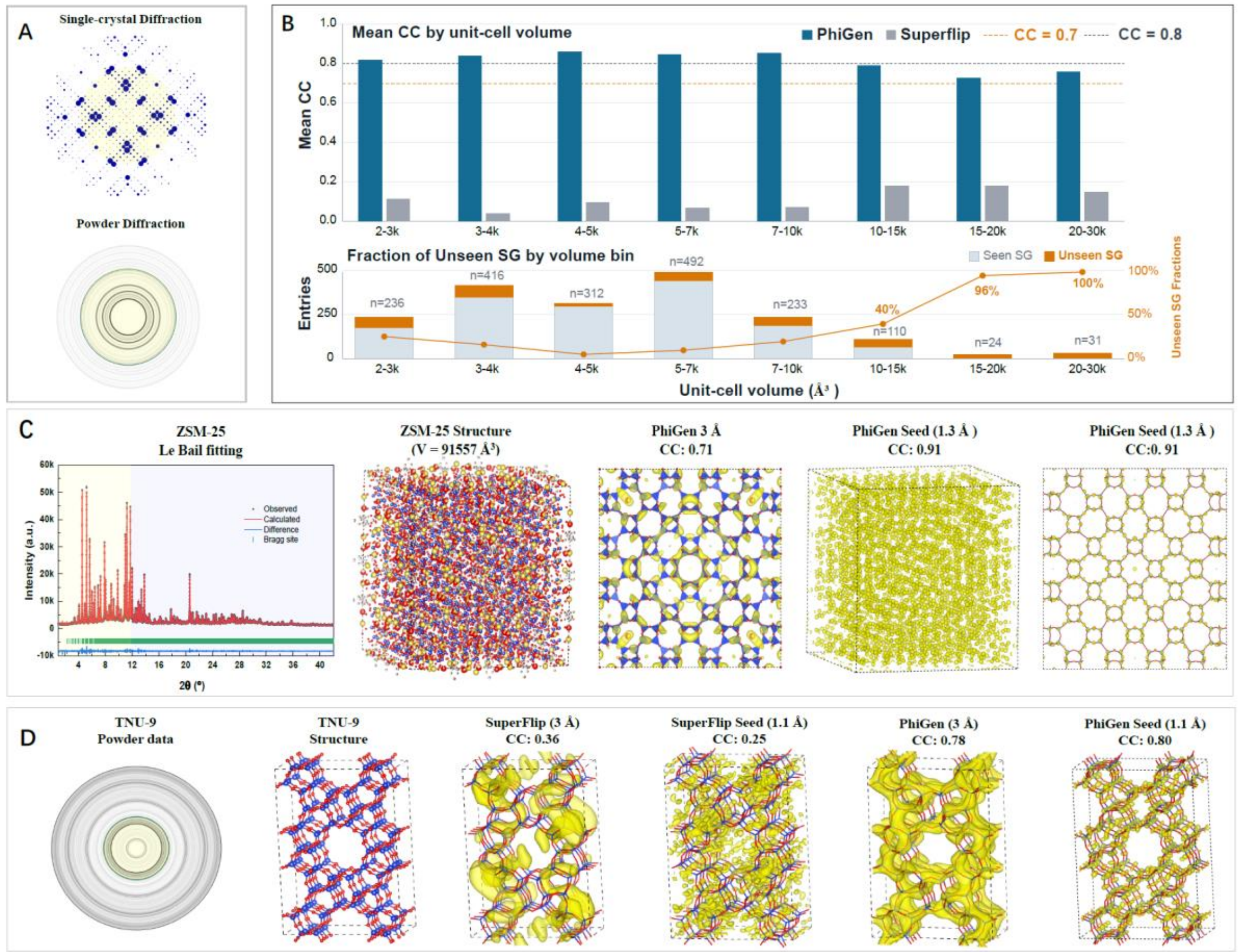

**Figure 5. PhiGen recovers zeolite frameworks from overlapped powder diffraction.** (A) Powder diffraction projects three-dimensional reciprocal-space intensities onto a one-dimensional profile, causing reflections with similar spacings to overlap. The yellow circle marks the 3 Å resolution limit for the simulated ZSM-25 single-crystal and powder data. (B) Benchmark performance across framework size and unseen symmetry for 1,854 simulated 3 Å powder zeolite datasets. Top: mean electron-density CC for PhiGen and Superflip across unit-cell-volume bins; dashed lines mark CC = 0.7 and CC = 0.8. Bottom: number of entries per bin and the fraction belonging to space groups absent from training. Across the full benchmark, PhiGen recovered framework-level maps for 84.2% of entries, compared with 1.0% for Superflip. In the 20,000–30,000 Å³ bin, PhiGen recovered 64.5% of structures, whereas Superflip recovered none. (C) Experimental ZSM-25 powder data structure solution. Structure-free Le Bail fitting yielded amplitudes to 3 Å (yellow) and 1.3 Å; 98.6% of 1.3 Å reflections belonged to overlap groups. A PhiGen 3 Å phase seed (CC = 0.71) was extended against the 1.3 Å data using Superflip, yielding a final map with CC = 0.91 that resolved all 16 independent tetrahedral sites and 80% framework oxygen sites. (D) Experimental TNU-9 powder data. A PhiGen 3 Å phase seed (CC = 0.78) was extended to 1.16 Å, producing a framework map with CC = 0.76 in which individual framework atoms were resolved. Superflip produced no solved extension; the best final CC was 0.25.

# Acknowledgments

The AI-driven experiments, simulations and model training were performed on the robotic AI-Scientist platform of Chinese Academy of Science. This work is financially supported by the National Natural Science Foundation of China (Grant No. 52272268, S.F.J.), (Grant No. 62376276, W.B.H.), (Grant No. 61925601, Y.L.), (Grant No. 12188101, H.M.W.), Beijing Nova Program (Grant No. 20230484278, W.B.H.), the Fundamental Research Funds for the Central Universities, and the Research Funds of Renmin University of China (Grant No. 23XNKJ19, W.B.H.), the National Key Research and Development Program of China (Grant No. 2022YFA1403800, H.M.W.), the New Cornerstone Science Foundation through the XPLORER PRIZE (H.M.W.).

## Author contributions:

S.F.J. and W.B.H. conceived the idea. Q.L. and R.J. designed the PhiGen model with physical constrains. L.M.W. and C.C. revised the code and some pictures. T.N.Z., B.T.W., Q.L.L., Z.L.P., M.N.H., Y.P.Y., L.Y., W.D., M.S., L.B. participated in the discussions and gave useful suggestions. S.F.J., Q.L. and R.J. wrote the manuscript and the supporting information. W.B.H., X.L.C., and L.M.W. revised and polished the manuscript. X.L.C., H.M.W., and Y.L. supervised the project. All authors discussed the results and commented on the final version of the paper.

## Competing interests:

There are no competing interests to declare.

## Data and materials availability:

Codes and datasets are available once our paper is published.